\documentclass[reprint, showpacs, superscriptaddress, aps, prl, twocolumn, 10pt]{revtex4-1}
\usepackage{graphicx}
\usepackage{color}
\usepackage{amssymb}
\usepackage{amsmath}
\usepackage{tabularx}
\usepackage{bm}
\usepackage{hyperref}
\usepackage{ltxtable}
\usepackage{natbib} 
\usepackage{longtable}
\usepackage{enumitem}
\usepackage{float}
\usepackage{mathtools}

\newcommand{\be}{\begin{equation}}
\newcommand{\ee}{\end{equation}}
\newcommand{\bse}{\begin{subequations}}
	\newcommand{\ese}{\end{subequations}}
\newcommand{\bea}{\begin{eqnarray}}
\newcommand{\eea}{\end{eqnarray}}
\newcommand{\ba}{\begin{array}}
	\newcommand{\ea}{\end{array}}

\usepackage[usenames,dvipsnames]{xcolor}
\hypersetup{
colorlinks,
citecolor=blue,
filecolor=blue,
linkcolor=blue,
urlcolor=blue}

\begin{document}

\title{Critical Prandtl number for Heat Transfer Enhancement in Rotating Convection}
\author{Mohammad Anas}
\affiliation{Department of Mechanical Engineering, Indian Institute of Technology, Kanpur 208016, India}
\author{Pranav Joshi}
\affiliation{Department of Mechanical Engineering, Indian Institute of Technology, Kanpur 208016, India}


\begin{abstract}
Rotation, which stabilizes flow, can enhance the heat transfer in Rayleigh-B\'enard convection (RBC) through Ekman pumping. In this Letter, we present the results of our direct numerical simulations of rotating RBC, providing a comprehensive analysis of this heat transfer enhancement relative to non-rotating RBC in the parameter space of Rayleigh number ($Ra$), Prandtl number ($Pr$), and Taylor number ($Ta$). We show that for a given $Ra$, there exists a critical Prandtl number ($Pr_{cr}$) below which no significant heat transfer enhancement occurs at any rotation rate, and an optimal Prandtl number ($Pr_{opt}$) at which maximum heat transfer enhancement occurs at an optimal rotation rate ($Ta_{opt}$). Notably, $Pr_{cr}$, $Pr_{opt}$, $Ta_{opt}$, and the maximum heat transfer enhancement all increase with increasing $Ra$. We also demonstrate a significant heat transfer enhancement up to $Ra=2\times 10^{10}$ and predict that the enhancement would become even more pronounced at higher $Ra$, provided $Pr$ is also increased commensurately.
\end{abstract}

\maketitle
Thermal convection under the influence of background rotation manifests in various geophysical and astrophysical flows, such as flows occurring within the Earth’s atmosphere, oceans, and outer core~\cite{Gill:book, Glatzmaier:Nature1995, Marshall:RGP1999}, gaseous planets like Jupiter~\cite{Ingersoll:Science1990, Heimpel:Nature2005}, and solar interiors~\cite{Spiegel:ARAA1971}. Rotation, which introduces the Coriolis force into the system, significantly affects the characteristics of these flows, including heat and momentum transfer~\cite{Chandrasekhar:book:Instability, Rossby:JFM1969}. The canonical model to study the behavior of such systems is rotating Rayleigh-B\'enard convection (RBC), in which fluid motion occurs between a hot plate (at the bottom) and a cold plate (at the top) as a consequence of the thermal buoyancy while the system rotates along an axis parallel to the gravity~\cite{Chandrasekhar:book:Instability}. 

Rotating RBC is primarily governed by three dimensionless parameters: the Rayleigh number ($Ra$), which represents the strength of the buoyancy force over the dissipative forces, the Prandtl number ($Pr$), which represents the ratio of the momentum diffusivity to thermal diffusivity, and the Taylor number ($Ta$), which represents the strength of the Coriolis force relative to the viscous force. To characterize the relative strength of convection over rotation, convective Rossby number ($Ro=\sqrt{Ra/TaPr}$) is commonly used. When $Ro\gg 1$, the buoyancy force dominates over the Coriolis force, and the heat transfer characteristics of rotating RBC systems are similar to those of corresponding non-rotating RBC~\cite{King:Nature2008, King:JFM2012, Stevens:PRL2013}. On the other hand, when $Ro\ll 1$, rotation becomes dominant and the heat transfer in rotating RBC, as compared to that of non-rotating case, is severely suppressed. Such rotating RBC system exhibits similarities to geostrophic flow, which is characterized by a force balance between pressure gradient and the Coriolis force~\cite{King:Nature2008, Kunnen:JT2021, Ecke:ARFM2023}.

Rotation, which suppresses the intensity of flow, enhances the heat transfer in RBC for a certain range of $Ra$, $Pr$, and $Ta$~\cite{Rossby:JFM1969, Zhong:JFM1993, Liu:PRL1997, Kunnen:PRE2006, Stevens:EJMFB2013, Ping:PRL2015, Zhong:PRL2009RotatingRBC, Stevens:PRL2013, Yang:PRF2020, Vorobieff:JFM2002, Stevens:PRL2009, Stevens:NJP2010, Stevens:POF2010, Stevens:PRE2011, Joshi:JFM2017, Weiss:PRL2010, Weiss:PRE2016, Chong:PRL2017}. This heat transfer enhancement in rotating RBC as compared to the non-rotating case is ascribed to Ekman pumping. Rotation generates columnar vortices aligned with the rotation axis in the flow, which in turn induce a secondary motion (parallel to the rotation) within the viscous boundary layer~\cite{Davidson:book:TurbulenceRotating}. This secondary motion facilitates the transport of hot fluid (at the bottom plate) and cold fluid (at the top plate) from the thermal boundary layers, leading to this enhancement in the heat transfer~\cite{Julien:JFM1996, Zhong:PRL2009RotatingRBC, Ecke:ARFM2023}.

Although prior studies have reported the effect of $Ra$, $Pr$, and $Ta$ (or $1/Ro$) on the heat transfer enhancement in rotating RBC~\cite{Zhong:PRL2009RotatingRBC, Stevens:NJP2010, Stevens:EJMFB2013, Yang:PRF2020, Hartmann:PRF2023}, a clear understanding of the possible enhancement as a function of these parameters has been missing. In this Letter, we provide a systematic comprehensive analysis of the heat transfer enhancement in rotating RBC in the $Ra$, $Pr$, $Ta$ parameter space. We explore a very wide range of Prandtl numbers, including very high $Pr$ ($\sim 1000$) that have not been studied earlier for rotating convection, to uncover the existence of a ‘critical’ Prandtl number, $Pr_{cr}$. We show that for each $Ra$ (at least within the range of $Ra=2\times 10^4-2\times 10^{10}$ explored in the present work), a significant heat transfer enhancement will occur only if the Prandtl number is greater than $Pr_{cr}$ that increases with increasing $Ra$. For a given $Ra$, we also provide a precise definition of the optimal Prandtl number $Pr_{opt}$ at which the maximum heat transfer enhancement occurs at an optimal rotation rate, $Ta_{opt}$, and show that $Pr_{opt}$ and $Ta_{opt}$ also increase with increasing $Ra$. Furthermore, we demonstrate a significant heat transfer enhancement up to $Ra=2\times 10^{10}$ and predict even higher enhancement at higher $Ra$.

For this study, we perform direct numerical simulations (DNS) of rotating RBC for a wide range of parameters: $Ra=g\beta\Delta H^3 /(\nu\kappa)=2\times 10^4-2\times 10^{10}$, $Pr=\nu/\kappa=1-1000$, and $Ta=4\Omega^2H^4/\nu^2=0-2\times 10^{12}$, and measure the heat transfer in terms of the Nusselt number $Nu=qH/(\lambda\Delta)$ . Here, $g$ is the acceleration due to gravity, $\beta$ is the thermal expansion coefficient, $\Delta$ is the temperature difference between the hot and cold plates, $H$ is the separation between the plates, $\nu$ is the kinematic viscosity, $\kappa$ is the thermal diffusivity, $\Omega$ is the system's rotation rate, $\lambda$ is the thermal conductivity of the fluid, and $q$ is the heat flux from the hot to cold plates. We perform simulations in a horizontally periodic rectangular domain of size $L\times L\times H$ ($L\times L$ in the horizontal directions) employing isothermal and no-slip (and impenetrable) boundary conditions at the top (cold) and bottom (hot) plates. For the simulations of non-rotating RBC ($Ta=0$) at moderate $Ra$, we use large aspect ratio ($\Gamma=L/H$) to avoid the effect of confinement on the Nusselt number~\cite{Huang:PRL2013}: $\Gamma=8$ for $Ra=2\times 10^4-10^6$ and $\Gamma=4$ for $Ra=10^7-10^8$. Considering the high computational cost at large $\Gamma$ for high $Ra$, we use $\Gamma=1$ for $Ra=5\times 10^8-2.3\times 10^9$ and $\Gamma= 0.5$ for $Ra=10^{10}$. To obtain $Nu$ at $Ra=2\times 10^{10}$ for non-rotating RBC, we fit the available $Nu$ data in the range $Ra=10^6-10^{10}$ to a power law and estimate the following values: $Nu\approx 0.114Ra^{0.304}\approx 154$ for $Pr=20$, $Nu\approx 0.131Ra^{0.298}\approx 154$ for $Pr=50$, and $Nu\approx 0.122Ra^{0.301}\approx 154$ for $Pr=100$.

Since the horizontal length scale of the flow in rotating convection, $\ell$, decreases with $Ta$, as $\ell\approx 0.9Ta^{-1/6}H$~\cite{King:JFM2013}, we use relatively lower aspect ratios (half or one-fourth of those for the corresponding non-rotating RBC cases) for some simulations of rotating RBC. In all simulations of rotating RBC, we ensure that $\ell/L<1/10$ (and also $\ell_c/L\lesssim 1/8$) which is sufficient to mitigate the effect of confinement on $Nu$~\cite{Kunnen:JFM2016, Julien:JFM1996}. Here, $\ell_c\approx 2.4 Ta^{-1/6}H$ is the horizontal length scale that develops at the onset of convection in rotating RBC for $Pr\gtrsim 0.68$~\cite{Chandrasekhar:book:Instability, Ecke:ARFM2023}. For more details about the simulations and the solver used in this study, please refer to the Supplementary Material~\cite{SM:ThisPaper}.

\begin{figure}

	\includegraphics[width=0.45\textwidth]{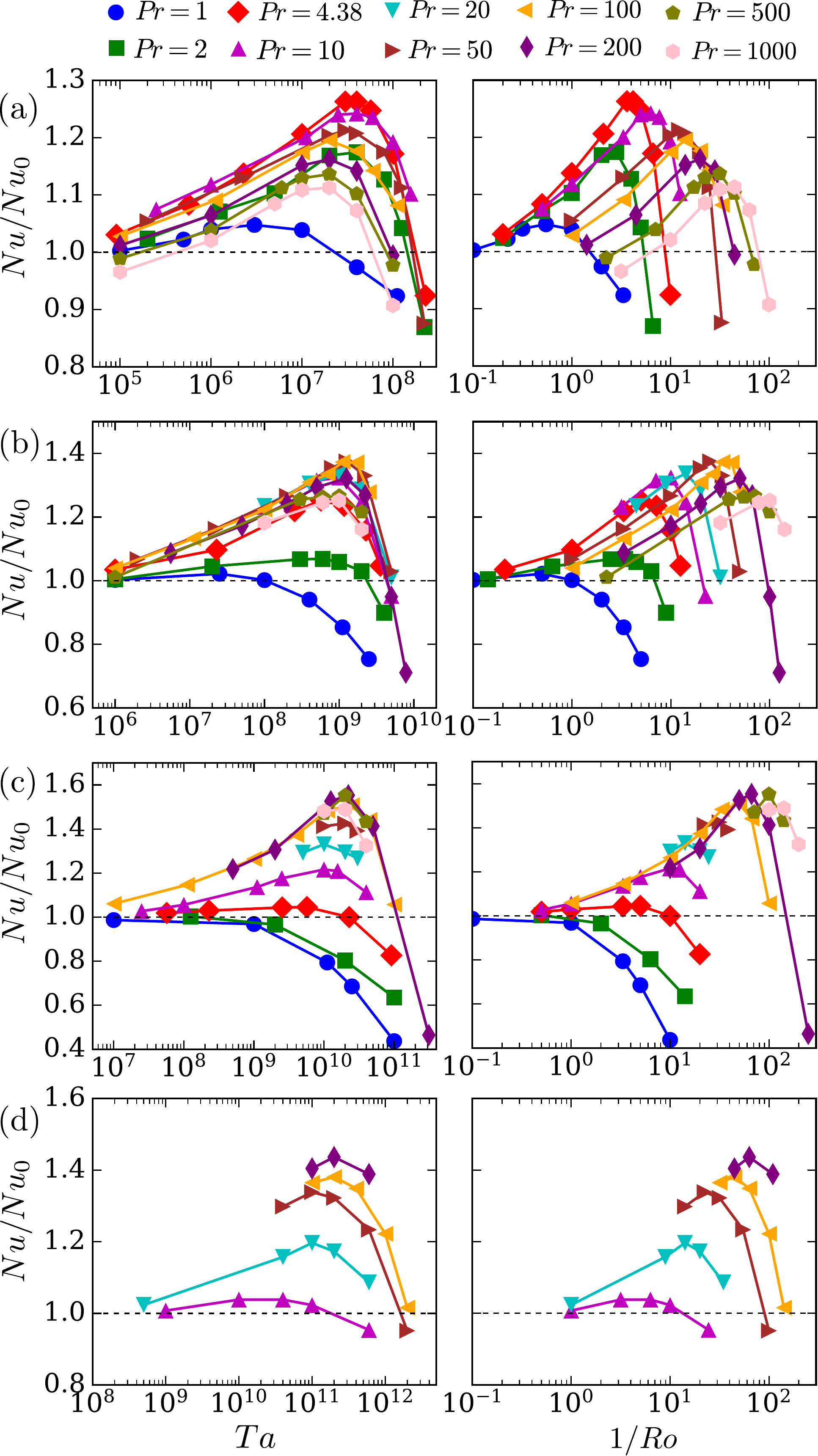}
	\caption{Variation of the normalized Nusselt number $Nu/Nu_0$ with Taylor number $Ta$ (left) and $1/Ro$ (right) for various $Pr$ at (a) $Ra=10^7$, (b) $Ra=10^8$, (c) $Ra=10^9$, and (d) $Ra=10^{10}$. Solid lines are used to connect data points, aiding visual interpretation.} 
	\label{fig:Nu_Ta_Ro}
\end{figure}

In Fig.~\ref{fig:Nu_Ta_Ro}, we show the variation of the normalized Nusselt number $Nu/Nu_0$ ($Nu_0$ is the Nusselt number for the non-rotating case) with the Taylor number, $Ta$, and the inverse of the Rossby number, $1/Ro$, for $Pr=1-1000$ at $Ra=[10^7, 10^8, 10^9, 10^{10}]$. We observe that when $Pr$ is not too small, $Nu/Nu_0$ first increases and then decreases as the rotation rate is increased. For each $Ra$, the maximum enhancement in the heat transfer as compared to the non-rotating case occurs in a certain range of $Pr$ and rotation rate. This maximum enhancement increases with increasing $Ra$ and can reach up to approximately $25\%$, $40\%$, and $55\%$ for $Ra=10^7$, $Ra=10^8$, and $Ra=10^9$, respectively. Note that we observe more than $40\%$ enhancement in heat transfer at $Ra=10^{10}$, which will be discussed later in greater detail.

Interestingly, we observe that the Taylor number $Ta$ serves as a better parameter than $1/Ro$ in representing the optimal rotation rate at which the maximum enhancement occurs (see Fig.~\ref{fig:Nu_Ta_Ro}). Unlike the optimal rotation rate represented in terms of the inverse of Rossby number, ($1/Ro_{opt}$), the optimal Taylor number, $Ta_{opt}$, is nearly independent of $Pr$ when a significant enhancement is observed at a given $Ra$. Most earlier studies used $1/Ro_{opt}$ to represent the optimal rotation rate but found $1/Ro_{opt}$ to be strongly dependent on $Pr$~\cite{Zhong:PRL2009RotatingRBC, Stevens:NJP2010, Yang:PRF2020}. Since the heat transfer enhancement due to rotation is largely controlled by the dynamics of the Ekman boundary layer~\cite{Stellmach:PRL2014}, the thickness of which depends only on the Taylor number (which represents the ratio of Coriolis to viscous forces), $Ta$ can be expected to represent better the heat transfer enhancement than $1/Ro$ (which represents the ratio of Coriolis to buoyancy forces) at moderate and high rotation rates. This finding is also in line with the hypothesis of ~\citet{King:Nature2008} that the boundary layer controls the rotation-dominated regime in rotating RBC, rather than the balance between the buoyancy and Coriolis forces. Nonetheless, the beginning of the rotation-affected regime, i.e., the rotation rate at which $Nu/Nu_0$ starts deviating from $1$, is better represented by $1/Ro$ than by $Ta$, as seen from our results. Specifically, for $Ra = 10^7$ and $10^8$, the rotation-affected regime begins at $1/Ro\approx 0.2$, while for $Ra=10^9$, it begins at $1/Ro\approx 0.4$. These values are consistent with the findings of~\citet{Stevens:PRL2009}.

\begin{figure}
	\includegraphics[width=0.38\textwidth]{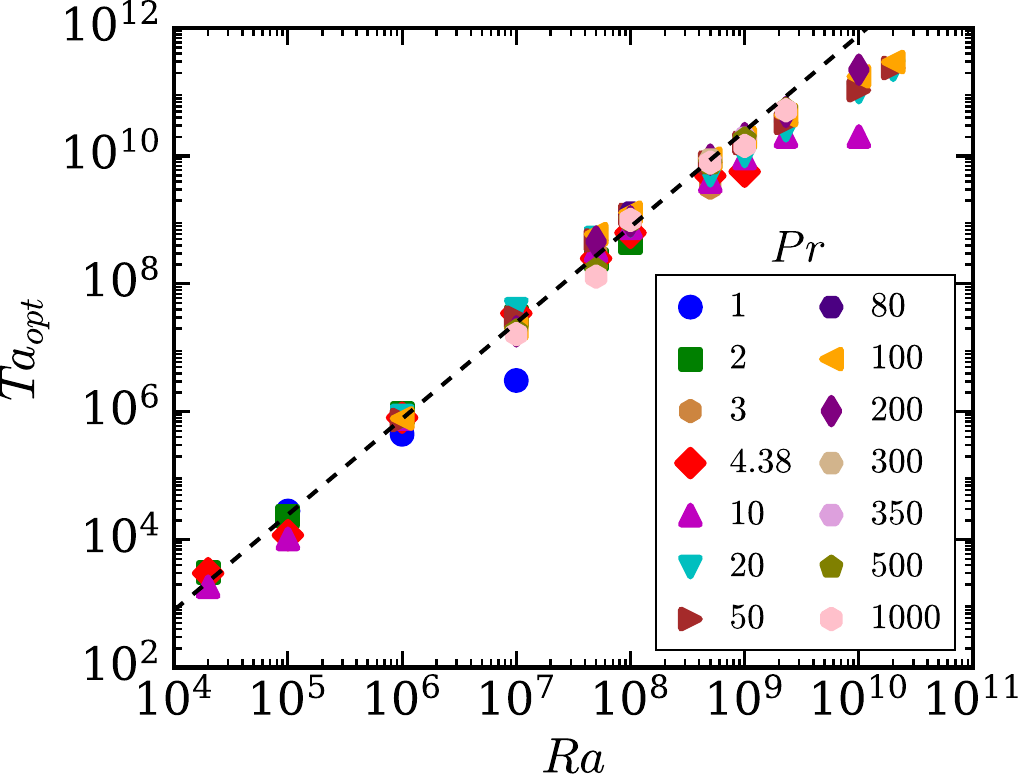}
	\caption{Variation of the optimal Taylor number $Ta_{opt}$ with $Ra$ for various $Pr$. Dashed line represents $Ta_{opt}=0.02Ta_{cs}$, where $Ta_{cs}=(Ra/8.7)^{1.5}$.} 
	\label{fig:Taopt_Ra}
\end{figure}

In Fig.~\ref{fig:Taopt_Ra}, we show the variation of $Ta_{opt}$ with $Ra$ for various Prandtl numbers. As discussed earlier, we observe that at a given $Ra$, the optimal Taylor number does not vary significantly with $Pr$. Also, $Ta_{opt}$ follows a power law close to $Ta_{opt}\propto Ra^{1.5}$ up to a certain $Ra$ and this limiting $Ra$ for the power law seems to increase with increasing $Pr$. Note that the Taylor number at which rotating RBC transitions from a convective state to a conductive state ($Ta_{cs}$) also follows the scaling $Ta_{cs}\propto Ra^{1.5}$ for $Pr\gtrsim 0.68$ and $Ta\gg 0$~\cite{Chandrasekhar:book:Instability, Ecke:ARFM2023}. Similar to $Ta_{cs}$, it is plausible that $Ta_{opt}$ also signifies a regime transition in rotating RBC which occurs at $Ta_{opt}\sim 0.02Ta_{cs}$ (see Fig.~\ref{fig:Taopt_Ra}).

\begin{figure}
	\includegraphics[width=0.45\textwidth]{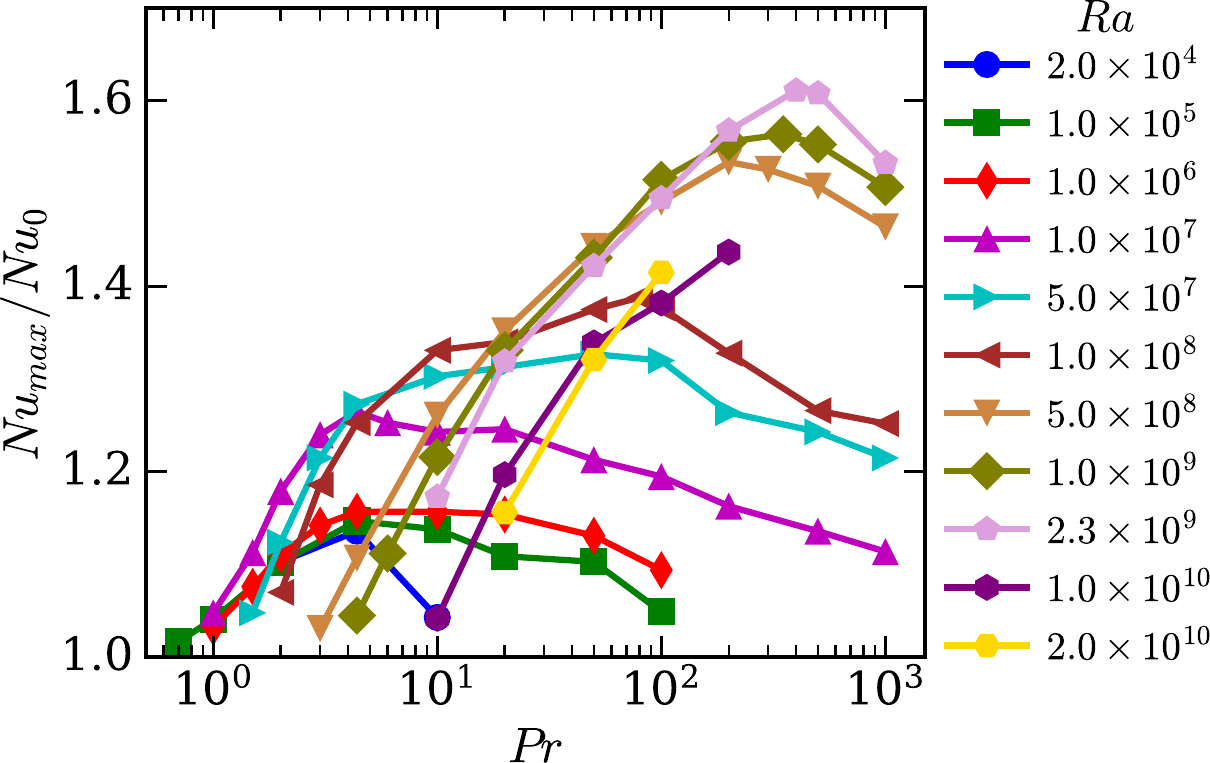}
	\caption{Variation of the normalized maximum Nusselt number $Nu_{max}/Nu_0$ with $Pr$ for various $Ra$.} 
	\label{fig:Nuopt_Pr}
\end{figure}

\begin{figure}
	\includegraphics[width=0.46\textwidth]{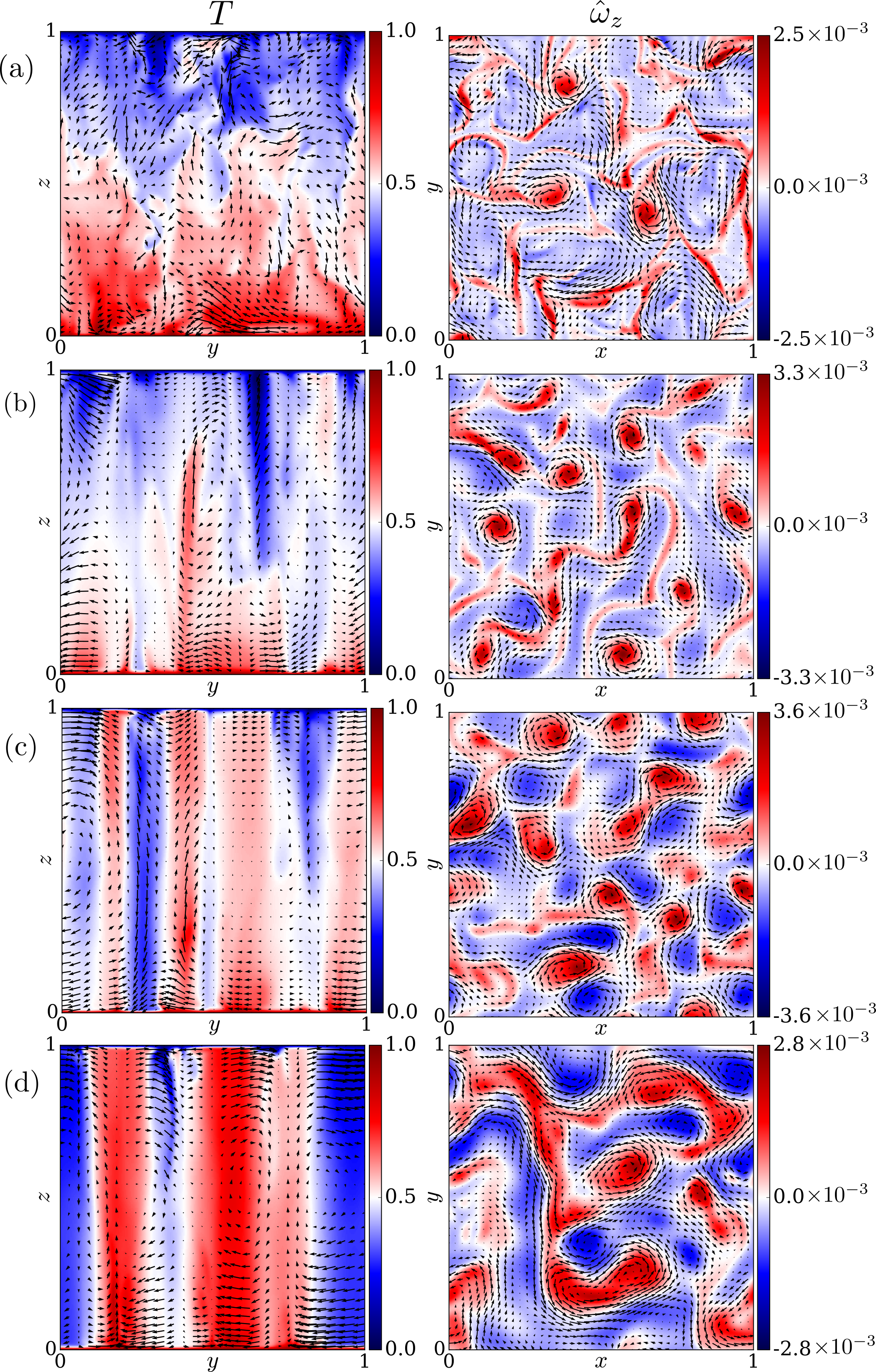}
	\caption{Instantaneous contours of the normalized temperature field $T$ in the $yz$-plane at $x=L/2$ (left) and the normalized $z$-vorticity field, $\hat\omega_z=\sqrt{(Pr/Ra)}\omega_z$, in the $xy$-plane at $z\approx 1.2\delta_u$ (right) with superimposed velocity vectors for (a) $Pr=1$, (b) $Pr=4.38$, (c) $Pr=100$, and (d) $Pr=1000$ at $Ra=10^8$ and $Ta\approx Ta_{opt}$. Note that the rotation axis is along $z$-direction and for $Ra=10^8$, $Pr_{cr}\approx 2$ and $Pr_{opt}\sim 100$.} 
	\label{fig:contourPlots}
\end{figure}

In Fig.~\ref{fig:Nuopt_Pr}, we show the variation of the normalized maximum Nusselt number, $Nu_{max}/Nu_0$, with $Pr$ for $Ra=2\times 10^4-2\times 10^{10}$. At any given $Ra$ and $Pr$, $Nu_{max}$, by definition, corresponds to the Nusselt number at the optimal Taylor number for that $Ra$ and $Pr$. The maximum heat transfer enhancement represented by $Nu_{max}/Nu_0$ for each $Ra$ increases with $Pr$ up to a certain $Pr$ and then decreases. Note that for each $Ra$ there exists a Prandtl number (obtained by extrapolating the data for each $Ra$ to $Nu_{max}/Nu_0=1$) below which there will be no (or marginal) heat transfer enhancement at any rotation rate. We call this $Pr$ the critical Prandtl number $Pr_{cr}$. In rotating convection, columnar vortical structures are known to play an important role in the heat transfer by transporting the temperature anomaly from the hot (cold) wall to the cold (hot) wall~\cite{Zhong:PRL2009RotatingRBC, Julien:JFM1996}. However, for $Pr<Pr_{cr}$, it is likely that the lateral diffusion of the heat/temperature anomaly away from the vortex columns restricts their ability to transport heat between the top and bottom walls~\cite{Stevens:NJP2010, Hartmann:PRF2023}: see the contours of the temperature field for $Pr=1$ in Fig.~\ref{fig:contourPlots}, which shows the instantaneous flow structures for various $Pr$ at $Ra=10^8$ and $Ta\approx 10^9$. Note that for $Ra=10^8$,  $Pr_{cr}\approx 2$, $Pr_{opt}\sim 100$, and $Ta_{opt}\sim 10^9$. As $Pr$ increases, this effect is expected to weaken {(e.g., see the contours of the temperature field in Fig.~\ref{fig:contourPlots}(b) and \ref{fig:contourPlots}(c))}, and so the heat transfer enhancement increases. However, the heat transfer enhancement decreases again at very large $Pr$. 
At any $Ra$, we define the Prandtl number at which $Nu_{max}/Nu_0$ reaches its maximum as the optimal Prandtl number $Pr_{opt}$ for that $Ra$. Some studies (e.g., \citet{Stevens:NJP2010}), comparing the heat transfer at a constant $Ro$, have proposed that at large $Pr$, the Ekman boundary layer ($\delta_u$) is significantly thicker than the thermal boundary layer ($\delta_\theta$); consequently, the columnar vortices do not reach the thermal boundary layer and the fluid entering them is not as hot (or as cold), leading to a decrease in the heat transfer enhancement at large $Pr$. However, we observe that the maximum enhancement (which occurs at $Ta_{opt}$) decreases at large Prandtl number even though $1.3\lesssim \delta_u/\delta_\theta\lesssim 1.5$ for $Pr>Pr_{opt}$ (see Supplementary Material~\cite{SM:ThisPaper}). 

Currently, we lack a concrete explanation for the existence of $Pr_{opt}$. However, to gain a preliminary understanding of the effect of $Pr$ on Ekman pumping and, consequently, on the heat transfer enhancement, we present the contours of the vorticity component in the direction of rotation ($z$-vorticity) along with the contours of the temperature fields in Fig.~\ref{fig:contourPlots}. Note that the normalized $z$-vorticity, $\hat\omega_z=\sqrt{(Pr/Ra)}\omega_z$, serves as a proxy for the strength of Ekman pumping~ \cite{Davidson:book:TurbulenceRotating}. It is clear that similar to the heat transfer enhancement, the strength of Ekman pumping (represented by $\hat\omega_z$) reaches a maximum at $Pr\sim Pr_{opt}$. Importantly, rotation seems to change profoundly the effect of $Pr$ on convection. At high $Pr$, a significant difference exists between the scales of the temperature and vorticity fields in non-rotating convection~\cite{Silano:JFM2010} (also see FIG. S5 in Supplementary Material~\cite{SM:ThisPaper}). However, as evident in Fig.~\ref{fig:contourPlots}(d) and FIG. S5 of ~\cite{SM:ThisPaper}, in rotating convection the temperature field maintains a high degree of coherence with $z$-vorticity. Consequently, the lateral spatial scales of the temperature field exhibit a substantial increase as $Pr$ increases beyond $Pr_{opt}$ and are likely to be associated with a decrease in the heat transfer enhancement in comparison to non-rotating convection.

Now, we make an important observation. Our results show a significant heat transfer enhancement compared to non-rotating RBC at $Ra=10^{10}$ ($\approx 43\%$ for $Pr=200$) and $Ra=2\times 10^{10}$ ($\approx 41\%$ for $Pr=100$), and the trends (see Fig.~\ref{fig:Nuopt_Pr}) suggest an even more pronounced enhancement for higher $Pr$ (up to the respective $Pr_{opt}$). Earlier, \citet{Weiss:PRE2016} also reported the heat transfer enhancement at $Ra\sim 10^{10}$ for $Pr=4.38-35.5$. However, the enhancement observed in their study, ($\approx 4\%-17\%$), is significantly lower than that in the present study because of their lower $Pr$ (closer to $Pr_{cr}$) and, in some cases, lower $Ta$ ($<Ta_{opt}$). The present results predict that a significant heat transfer enhancement in rotating RBC is possible at high $Ra$, provided $Pr$ is substantially higher (closer to $Pr_{opt}$) than $Pr_{cr}$. Since most earlier studies commonly employed air ($Pr\sim 1$) or water ($Pr\approx 4-7$) as the working fluid, for example, \citet{Niemela:JFM2010} ($Pr=0.7-5.9$), \citet{Stellmach:PRL2014} ($Pr\approx 1-7$), \citet{Kunnen:JFM2016} ($Pr=1$), \citet{Ecke:PRL2014} ($Pr=0.7$), and \citet{Hartmann:PRF2023} ($Pr = 4.38$ and $6.4$), they failed to observe any (or significant) heat transfer enhancement at $Ra\gtrsim 10^{10}$, for which $Pr_{cr}\gtrsim 10$. Note that we also observe a significant heat transfer enhancement ($>10\%$) at $Ra=2\times 10^4$ (using $\Gamma=8$), in agreement with \citet{Rossby:JFM1969}'s experimental results for $\Gamma\gtrsim 6$. 

\begin{figure}
	\includegraphics[width=0.35\textwidth]{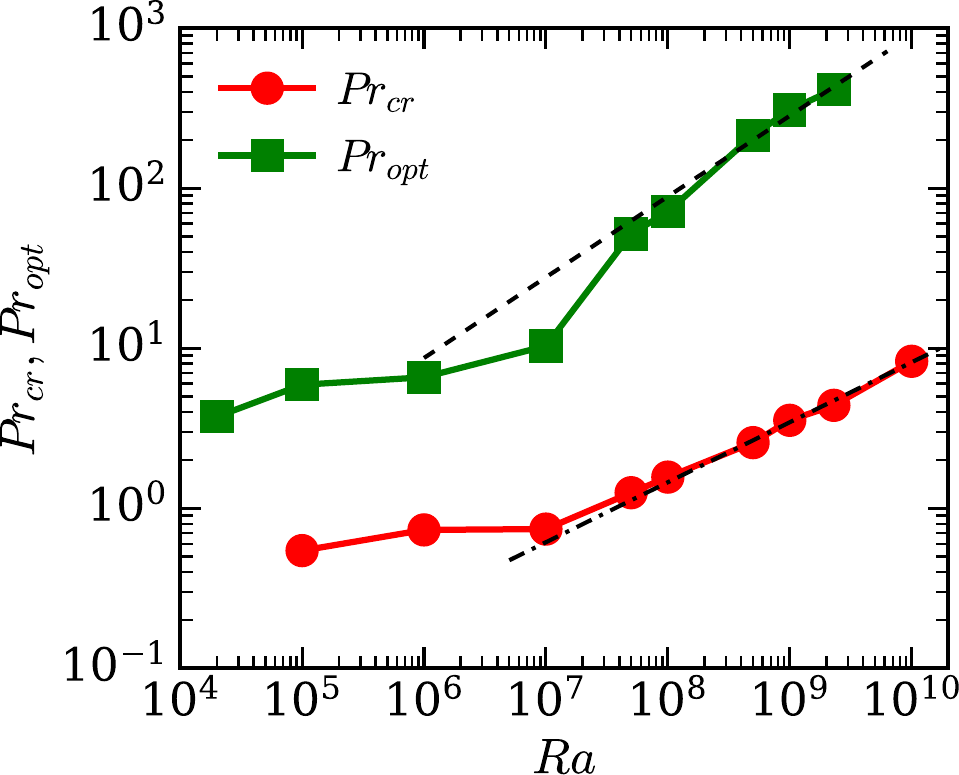}
	\caption{Variation of the critical Prandtl number $Pr_{cr}$  and the optimal Prandtl number $Pr_{opt}$ with $Ra$. Dot-dashed and dashed lines represent power-law fits $Pr_{cr}\approx 1.46\times 10^{-3}Ra^{0.375}$ in the range $Ra=10^7-10^{10}$ and $Pr_{opt}\approx 8.18\times 10^{-3}Ra^{0.504}$ in the range $Ra=10^6-2.3\times 10^9$, respectively.} 
	\label{fig:PrcrProptRa}
\end{figure}

In Fig.~\ref{fig:PrcrProptRa}, we show the variation of $Pr_{cr}$ and $Pr_{opt}$ with $Ra$. Interestingly, both $Pr_{cr}$ and $Pr_{opt}$ increase monotonically with $Ra$ and approximately follow power-laws: $Pr_{cr}\approx 1.46\times 10^{-3}Ra^{0.375}$ and $Pr_{opt}\approx 8.18\times 10^{-3}Ra^{0.504}$. Considering the high computational cost at large $Pr$ and large $Ra$, we do not perform simulations at $Pr>200$ and $Pr>100$ to find $Pr_{opt}$ for $Ra=10^{10}$ and $Ra=2\times10^{10}$, respectively, which are estimated to be $Pr_{opt}\sim 1000$ by the above power-law fit. As Rayleigh number increases, the turbulent diffusion of heat is also expected to become stronger. At low $Pr$, this higher turbulent diffusion will combine with the large molecular thermal diffusivity to further increase the lateral diffusion of heat in the bulk, and hence, will decrease the ability of the vortex columns to transport heat. Thus, a correspondingly larger $Pr$ may be necessary to counter this effect of the enhanced turbulent thermal diffusivity to register any enhancement in the heat transfer, i.e., $Pr_{cr}$ will increase with increasing $Ra$. On the other hand, as the buoyancy forcing increases with increasing $Ra$, the effects of large $Pr$ will diminish and the heat transfer enhancement can be sustained until larger Prandtl numbers, i.e. $Pr_{opt}$ also increases as $Ra$ is increased. 

\begin{figure}
	\includegraphics[width=0.35\textwidth]{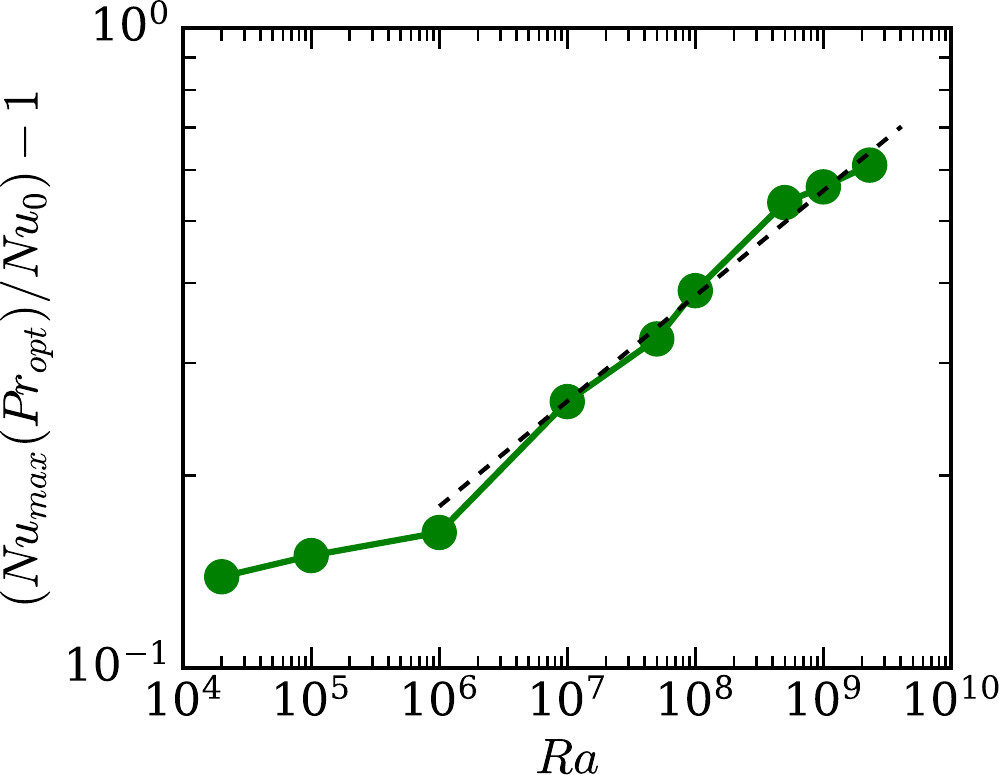}
	\caption{Variation of $(Nu_{max}(Pr_{opt})/Nu_0)-1$ with $Ra$. Dashed line represents power-law fit $Nu_{max}(Pr_{opt})/Nu_0\approx 1+0.0185Ra^{0.164}$ in the range $Ra=10^6-2.3\times 10^9$}. 
	\label{fig:NumaxRa}
\end{figure}

In Fig.~\ref{fig:NumaxRa}, we show the variation of $(Nu_{max}(Pr_{opt})/Nu_0)-1$ with $Ra$. Here, $Nu_{max}(Pr_{opt})$ is $Nu_{max}$ at $Pr=Pr_{opt}$. Similar to $Pr_{cr}$ and $Pr_{opt}$, $(Nu_{max}(Pr_{opt})/Nu_0)-1$ also increases with increasing $Ra$, and the trend can be fitted by a power-law $Nu_{max}(Pr_{opt})/Nu_0\approx 1+0.0185Ra^{0.164}$ in the range $Ra=10^6-2.3\times 10^9$. This relationship predicts that the maximum heat transfer enhancement increases with increasing $Ra$, provided the Prandtl number is also increased commensurately to the respective $Pr_{opt}$. In particular, the present results predict $Nu_{max}/Nu_0\approx 1.8$ (i.e., a maximum heat transfer enhancement of $80\%$) for $Ra=10^{10}$ at $Pr_{opt}\approx 900$, and even higher maximum enhancement at $Ra>10^{10}$ at higher $Pr$. As discussed earlier, $Pr_{opt}$ increases with $Ra$, i.e., the increase in the heat transfer enhancement with increasing $Pr$ can be sustained up to a larger $Pr$ at higher $Ra$. Thus, the maximum enhancement can also be expected to increase with $Ra$.

Note that, due to the effect of finite aspect ratio on $Nu$ in non-rotating simulations (mainly for $Ra\ge 5\times 10^8$ in this work)~\cite{Huang:PRL2013}, and owing to possible errors associated with the interpolation and extrapolation, there may be some uncertainty in the present values of $Pr_{cr}$, $Pr_{opt}$, and the maximum enhancement. However, this effect is expected to not alter any of the major findings of this study.

In conclusion, this work provides a clear picture of the heat transfer enhancement due to Ekman pumping in rotating Rayleigh-B\'enard convection (RBC) as a function of $Ra$, $Pr$, and $Ta$. At a given $Ra$, we demonstrate the existence of a critical Prandtl number below which no significant heat transfer enhancement can occur and an optimal Prandtl number at which the maximum enhancement is obtained. Both critical and optimal Prandtl numbers increase with increasing $Ra$. Importantly, our results show that the maximum enhancement also increases with $Ra$, provided $Pr$ is increased commensurately. At present, we do not know up to what $Ra$ these trends will persist. Simulations and experiments at significantly higher $Ra$ and $Pr$ than currently feasible may be required to answer this question.  
	
Note that, similar to rotation, lateral confinement and a vertical magnetic field also stabilize flow and can also enhance the heat transfer in RBC~\cite{Huang:PRL2013, Chong:PRL2015, Chong:PRL2017, Chong:JFM2019}. Perhaps a systematic study of the effects of lateral confinement and a vertical magnetic field on the heat transfer in RBC will also reveal the existence of critical and optimal Prandtl numbers for these problems. In closing, we want to emphasize that by systematically exploring the heat transfer over a wide range of parameters, we believe that our study establishes a benchmark for testing and developing new models for rotating RBC.
\\

For all the simulations related to this work, we gratefully acknowledge the support and the resources provided by Param Sanganak under the National Supercomputing Mission, Government of India at the Indian Institute of Technology, Kanpur. We also thank Prof. M. K. Verma for inspiring us to utilize GPUs for scientific computing, and also for providing resources for the testing of the solver and for running some simulations. Mohammad Anas thanks Roshan Samuel for his valuable assistance in the development of the solver, as well as Soumyadeep Chatterjee, Shadab Alam, and Manthan Verma for their useful discussions on the solver and this work.


%

\end{document}